\documentclass[final,5p,times,twocolumn]{elsarticle}
\usepackage{graphics}
\usepackage{amsmath}
\usepackage{color}
\usepackage{marvosym}
\usepackage{multirow} 
\usepackage{booktabs}
\usepackage{bm}
\usepackage{ulem}
\usepackage[switch]{lineno}
\usepackage{makecell}
\usepackage{url}
\journal{Information Fusion}
\newcolumntype{L}[1]{>{\raggedright\arraybackslash}p{#1}}
\newcolumntype{C}[1]{>{\centering\arraybackslash}p{#1}}
\newcolumntype{R}[1]{>{\raggedleft\arraybackslash}p{#1}}
\newcommand\clr[1]{{\color{black}{#1}}}

\newcommand{\settablefont}{\fontsize{6.5}{9.0}\selectfont}

\begin{document}
\begin{frontmatter}
\title{ODFormer: Semantic Fundus Image Segmentation Using Transformer for\\Optic Nerve Head Detection}
\author[inst11,inst12,inst13]{Jiayi Wang}
\author[inst2]{Yi-An Mao}
\author[inst3]{Xiaoyu Ma}
\author[inst11,inst12,inst13]{Sicen Guo}
\author[inst3]{Yuting Shao}
\author[inst3]{Xiao Lv}
\author[inst3]{Wenting Han}
\author[inst4]{Mark Christopher}
\author[inst4]{\\Linda M. Zangwill}
\author[inst3,inst5]{Yanlong Bi\textsuperscript{\Letter}}
\author[inst11,inst12,inst13]{Rui Fan\textsuperscript{\Letter}}

\affiliation[inst11]{organization={Machine Intelligence \& Autonomous Systems (MIAS) Group, the College of Electronics \& Information Engineering, Tongji University},
	city={Shanghai},
	postcode={201804}, 
	country={China}}
\affiliation[inst12]{organization={Shanghai Institute of Intelligent Science and Technology, Tongji University},
	city={Shanghai},
	postcode={201804}, 
	country={China}}
\affiliation[inst13]{organization={Shanghai Research Institute for Intelligent Autonomous Systems, the State Key Laboratory of Intelligent Autonomous Systems, and Frontiers Science Center for Intelligent Autonomous Systems, Tongji University},
	city={Shanghai},
	postcode={201210}, 
	country={China}}

\affiliation[inst2]{organization={School of Life Science, Shanghai University},
	addressline={99 Shangda Avenue, Baoshan District}, 
	city={Shanghai},
	postcode={200444}, 
	country={China}}

\affiliation[inst3]{organization={Department of Ophthalmology, Tongji Hospital, School of Medicine, Tongji University},
	city={Shanghai},
	postcode={200071},
	country={China}}    

\affiliation[inst4]{
	organization={Hamilton Glaucoma Center and Division of Ophthalmology Informatics and Data Science, Shiley Eye Institute, Viterbi Family Department of Ophthalmology, University of California, San Diego},
	city={San Diego},
	postcode={CA 92037}, 
	country={USA}
}

\affiliation[inst5]{organization={Tongji Eye Institute, School of Medicine, Tongji University, Shanghai, China},
	city={Shanghai},
	postcode={200071},
	country={China}}

\begin{abstract}
\clr{
Optic nerve head (ONH) detection has been a crucial area of study in ophthalmology for years. However, the significant discrepancy between fundus image datasets, each generated using a single type of fundus camera, poses challenges to the generalizability of ONH detection approaches developed based on semantic segmentation networks. Despite the numerous recent advancements in general-purpose semantic segmentation methods using convolutional neural networks (CNNs) and Transformers, there is currently a lack of benchmarks for these state-of-the-art (SoTA) networks specifically trained for ONH detection. Therefore, in this article, we make contributions from three key aspects: network design, the publication of a dataset, and the establishment of a comprehensive benchmark. Our newly developed ONH detection network, referred to as ODFormer, is based upon the Swin Transformer architecture and incorporates two novel components: a multi-scale context aggregator and a lightweight bidirectional feature recalibrator. Our published large-scale dataset, known as TongjiU-DROD, provides multi-resolution fundus images for each participant, captured using two distinct types of cameras. Our established benchmark involves three datasets: DRIONS-DB, DRISHTI-GS1, and TongjiU-DROD, created by researchers from different countries and containing fundus images captured from participants of diverse races and ages. Extensive experimental results demonstrate that our proposed ODFormer outperforms other state-of-the-art (SoTA) networks in terms of performance and generalizability. Our dataset and source code are publicly available at \url{mias.group/ODFormer}.
}
\end{abstract}
\begin{keyword}
\clr{Optic nerve head \sep fundus image \sep semantic segmentation \sep convolutional neural network \sep Transformer}
\end{keyword}

\end{frontmatter}

\section{Introduction}
\label{sec:introduction}

\subsection{Background}

\clr{Fundus images, which are photographs of the back of the eyes, are crucial for diagnosing not only retinal and ophthalmic conditions but also cerebrovascular and other systemic diseases \cite{wong2001retinal}. Among the various anatomical features visible in a fundus image, the optic nerve head (ONH) is of particular significance \cite{sinthanayothin1999automated}. The ONH generally presents as a bright yellowish area, slightly oval in shape, with blood vessels converging towards its center \cite{abdullah2016localization}. It consists of two distinct zones: (1) the central zone, or the cup, and (2) the peripheral zone, or neuroretinal rim \cite{cheng2013superpixel}. The ONH is critical in fundus image analysis as it serves as a reference point for various diagnostic tasks, including fovea location estimation \cite{rohrschneider2004determination} and cup-to-disc ratio calculation \cite{zhang2020intelligent}. The detection of the ONH is thus of considerable diagnostic importance \cite{fan2023tmi, li2001automatic, fan2022detecting, abdel2004detection, fan2023detecting}. It is a fundamental step in computer-aided systems that assist ophthalmologists in diagnosing conditions such as diabetic retinopathy \cite{bharkad2017automatic}. Moreover, accurate ONH detection can provide valuable diagnostic insights. For instance, the size of the ONH, derived from its detection, is commonly used to diagnose glaucoma by assessing the cup-to-disc ratio \cite{chrastek2005automated}. Achieving precise ONH detection is, therefore, crucial in ophthalmology, significantly impacting both diagnosis and treatment.}

\subsection{Motivation}
\clr{The rapid progress in deep learning technologies has spurred the development of numerous semantic segmentation approaches based on neural networks, which are powerful tools for accurate ONH detection.} Compared to traditional \clr{Convolutional Neural Networks (CNNs)}, the advantage of Transformers equipped with self-attention mechanisms lies in providing a more effective strategy for global context modeling \cite{li2023uniformer}. \clr{They have proven superior, especially in terms of generalizability on new, unseen datasets, compared to CNNs in various foundational computer vision tasks \cite{fan2023detecting}.} Several endeavors have been made to integrate the strengths of Transformer and CNN. For example, Polarformer \cite{feng2022polarformer} introduces a hybrid CNN-Transformer and polar transformation network, leveraging the Transformer's large receptive fields to incorporate global contextual information. Similarly, based on U-Net, UT-Net \cite{hussain2023ut} employs a Transformer to retain learned spatial gradient information and \clr{utilizes} this knowledge in subsequent stages, enabling improved identification of local minor gradient changes near the boundary region. \clr{Nevertheless, despite the impressive capabilities Transformers have demonstrated in vision tasks, their performance still has weaknesses due to the following three factors:} 
\clr{\begin{enumerate}
    \item The architecture often overlooks local spatial priors within each patch when dividing images into sequences of non-overlapping patches.
    \item Transformers struggle to explicitly extract structural information due to the use of absolute positional encodings.
    \item  The frequent use of upsampling to leverage the multi-scale features generated by the Transformer can lead to the loss of low-resolution details, posing significant challenges for dense prediction tasks such as detection and segmentation.
\end{enumerate}
}

\clr{Variations in fundus imaging technology lead to diverse morphological characteristics across different fundus image datasets, which poses a significant challenge for many ONH detection methods in achieving robust generalizability. For instance, portable fundus cameras provide adaptability and convenience in various settings but often yield images of varying quality compared to professional desktop cameras. While advanced cameras deliver high-quality images that are crucial for accurate disease diagnosis, their higher costs and larger sizes may limit accessibility and pose practical challenges in certain environments. Additionally, there is a scarcity of comprehensive fundus image datasets specifically designed to overcome this limitation. Most existing datasets are captured using only one type of fundus camera, which restricts the evaluation of performance and generalizability of segmentation algorithms. Furthermore, there is a notable absence of a dedicated benchmark for ONH detection using state-of-the-art (SoTA) semantic segmentation networks. Given these challenges, there is a compelling need to conduct a comprehensive comparison of SoTA networks, trained and validated across various datasets, to accurately assess the performance and generalizability of segmentation networks in ONH detection.}

\subsection{Novel Contributions}
\label{sec.novelties}

To address the above-mentioned limitations, 
we introduce \uline{\textbf{O}NH \textbf{D}etection Trans\textbf{former} (\textbf{ODFormer})}, a novel semantic segmentation network, developed specifically for ONH detection. We first present a Multi-Scale Context Aggregator (MSCA), which partitions the input image into patches while simultaneously enlarging the receptive field using a series of atrous convolutions with increasing atrous rates. Additionally, building upon the Swin Transformer \cite{liu2021swin} architecture, we design an encoder that improves the position encoding in self-attention calculations. 
\clr{This is achieved by integrating a relative position bias map extracted through an additional convolutional layer, obtaining more comprehensive relative and absolute position information.}
Regarding our ODFormer decoder, which is developed based on \clr{UPerNet} \cite{xiao2018unified}, a significant contribution lies in the Lightweight Bidirectional Feature Recalibrator (LBFR). It alleviates the loss of high-frequency spatial information resulting from both the upsampling operation and the fusion of feature maps within the pyramid structure. Moreover, we publish a large-scale, comprehensive fundus image dataset, referred to as \uline{\textbf{Tongji} \textbf{U}niversity \textbf{D}ual-\textbf{R}esolution \textbf{O}NH \textbf{D}etection (\textbf{TongjiU-DROD})} dataset. For each participant, fundus images were captured using two distinct cameras, and the semantic segmentation \clr{ground-truth annotations} for these images were manually annotated. We establish an ONH detection benchmark using our published TongjiU-DROD dataset, in conjunction with two publicly available datasets: DRIONS-DB \cite{carmona2008identification} and DRISHTI-GS1 \cite{sivaswamy2015comprehensive}. \clr{This benchmark enables both qualitative and quantitative comparisons of SoTA CNNs and Transformers for ONH detection. Extensive experimental results demonstrate} (1) the effectiveness of our designed modules within ODFormer, (2) its superior performance and generalizability across various datasets, and (3) the suitability of our curated dataset for training and evaluating ONH detection networks.

In summary, the contributions of this article are as follows:
\begin{enumerate}
    \item 
    \clr{A novel ONH detection network, referred to as ODFormer;}
    \item 
    \clr{A new dataset for ONH detection, captured using two different types of fundus cameras;}
    \item  
    A benchmark of SoTA semantic segmentation CNNs and Transformers trained on the three aforementioned datasets for ONH detection.
\end{enumerate}

\subsection{Article Structure}
\label{sec.paper_structure}
The remainder of this article is structured as follows: Sect. \ref{Literature Review} reviews the SoTA CNNs-based and Transformer-based semantic segmentation networks. Sect. \ref{Methodology} introduces our proposed ODFormer. Sect. \ref{sec.datasets} details our proposed TongjiU-DROD dataset. Sect. \ref{Experiment Results and Discussion} presents the experimental results and compares our \clr{network} with other SoTA networks. Finally, we conclude this article in the last section.

\section{Literature Review}
\label{Literature Review}

\subsection{\clr{Semantic Segmentation CNNs}}

\clr{
Fully Convolutional Network (FCN) \cite{long2015fully} is a pioneering work for end-to-end semantic segmentation. Unlike traditional CNNs, which often employ multiple fully connected layers after the final convolutional layer to transform a feature map into a fixed-length feature vector, FCNs can process input images of any size. It upsamples the feature map of the final layer using a deconvolution layer, allowing the output to revert to the original image size and generate pixel-level predictions while maintaining the spatial information of the original input. However, FCNs are somewhat limited in leveraging global scene category cues. The Pyramid Scene Parsing Network (PSPNet) \cite{zhao2017pyramid} addresses this limitation by performing spatial pooling at various grid scales, which has demonstrated outstanding performance on several semantic segmentation benchmarks. PSPNet combines a global pyramid pooling structure to expand pixel-level features with the extended FCN architecture \cite{zhou2016image} for semantic segmentation, without significantly increasing the computational expense in terms of the number of parameters. The Feature Pyramid Network (FPN) \cite{lin2017feature}, a general-purpose feature extractor, capitalizes on multi-level feature representations organized in a natural, pyramidal hierarchy. It introduces a minor computational overhead through a top-down architecture and lateral connections that fuse high-level semantic information with intermediate and low-level details. Furthermore, the Unified Perceptual Parsing Network (UPerNet) \cite{xiao2018unified} integrates the architecture and the Pyramid Pooling Module (PPM) from PSPNet \cite{zhao2017pyramid}, addressing the issue of a relatively small empirical receptive field in deep CNNs. 

Unlike PSPNet, DeepLabv3 \cite{florian2017rethinking} employs multiple parallel atrous convolutions, known as Atrous Spatial Pyramid Pooling (ASPP), with varying rates to capture contextual information across different scales. While this approach enables the final feature map to hold rich semantic information, it often lacks accurate object boundary details due to the use of pooling and striding operations in the network's backbone. To address this shortfall, DeepLabv3+ \cite{chen2018encoder} introduces a concise yet efficient decoder to the existing DeepLabv3 architecture. This modification significantly enhances segmentation results, especially along object boundaries, by refining segmentation details and aligning more closely with object contours. Both ASPP \cite{florian2017rethinking} and PPM \cite{zhao2017pyramid} utilize predefined convolutions with different dilation rates and multi-scale pooling operations. These methods are sensitive to the size of input images and the scale differences between images during the training and inference phases. The fixed weights, predefined dilation rates, and pooling grids often struggle to adapt to internal scale variations present in input images of arbitrary sizes and scales, which can lead to less optimal performance under varied imaging conditions.

The self-attention module enhances an individual element by aggregating features from a set of elements, with aggregation weights typically based on the embedded feature similarities between these elements. This mechanism is effective at capturing contexts and long-range dependencies. With the widespread integration of attention mechanisms into neural networks, the Non-local neural network (NonLocal) \cite{wang2018non} computes the attention mask by deriving a correlation matrix between every spatial point in the feature map. However, the computational demands and substantial GPU memory consumption of NonLocal limit its practical deployment in many real-world applications. To improve efficiency and reduce memory usage without sacrificing the overall performance, the Asymmetric Non-local Neural Network (ANN) \cite{zhu2019asymmetric} selectively samples representative points from feature maps. Meanwhile, the Global Context Network (GCNet) \cite{cao2019gcnet} simplifies the non-local block by computing a global attention map and sharing it across all locations, significantly reducing computational load. The Dual Attention Network (DANet) \cite{fu2019dual} encodes global context through a carefully designed self-attention mechanism that incorporates both spatial and channel attention. While DANet adapts weights to compute pair-wise similarity or learn pixel-wise attention maps, it tends to overlook the role of global guidance from the global information extractor. In contrast, APCNet considers global-guided local affinity to estimate the contribution degree of subregions from both local and global representations, leveraging multi-scale representation using a feature pyramid. The Dynamic Multi-scale Network (DMNet) \cite{he2019dynamic} learns variable-scale features with dynamic multi-scale filters that adaptively select important areas, enabling the model to better extract contextual information. This approach provides greater adaptability and flexibility, as each branch can capture distinct scale features relevant to the input image, offering enhanced capability in handling diverse scenarios.
}

\begin{figure*}[t]
\includegraphics[width=1\textwidth]{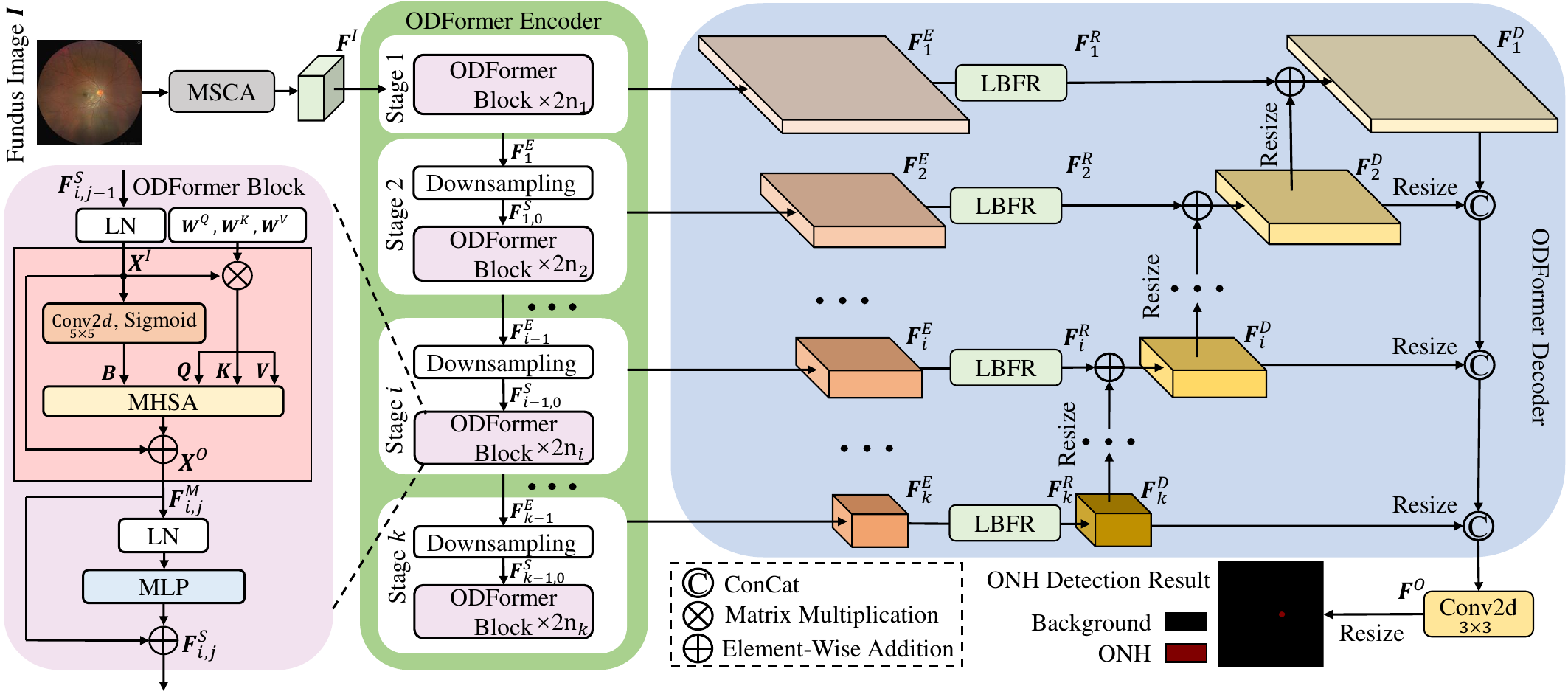}
\caption{An overview of our proposed ODFormer.}
\label{fig.architecture}
\end{figure*}

\subsection{\clr{Semantic Segmentation Transformers}}
\clr{
Self-attention is increasingly recognized as a crucial component in CNN architectures, particularly due to its ability to effectively scale with large receptive fields \cite{liu2024playing}. This building block is commonly applied atop networks to capture long-range interactions and enhance high-level semantic features essential for vision tasks \cite{han2022survey}. Several networks have refined the self-attention mechanism to maximize its benefits, achieving impressive results. For instance, DeFusionNET \cite{tang2020defusionnet} introduces a newly designed channel attention module that selectively emphasizes discriminative features, significantly enhancing the feature refinement process.

Self-attention also plays a pivotal role in Transformers \cite{vaswani2017attention}, which utilize this mechanism to capture long-range dependencies among tokens within a sentence. Moreover, Transformers are highly adept at parallelization, facilitating efficient training on large datasets \cite{li2024roadformer,li2024hapnet}. The success of Transformers in Natural Language Processing (NLP) has spurred the adoption of similar models in computer vision, particularly for semantic segmentation problems \cite{strudel2021segmenter}. Inspired by these successes, researchers have explored the potential of Transformers to learn useful representations from images. In the domain of semantic segmentation, the Object-Contextual Representation (OCR) \cite{yuan2020object} utilizes Transformers to enhance pixel representations by associating each pixel with the representation of its corresponding object class. This technique significantly improves the accuracy of pixel classification. Further demonstrating the capabilities of Transformers in computer vision, networks such as SegFormer and Twins have made notable contributions. SegFormer \cite{xie2021segformer} combines the efficiency of Transformers with lightweight multilayer perceptron (MLP) decoders, creating a simple yet powerful semantic segmentation framework. Twins \cite{chu2021twins} introduces two innovative vision Transformer (ViT) backbones, Twins-PCPVT and Twins-SVT, showing that a combination of local and global attention can yield impressive results, albeit with increased computational demands. Additionally, the Swin Transformer \cite{liu2021swin} introduces a variant of ViT \cite{dosovitskiy2020image} which has a hierarchical architecture. This model utilizes shifted windows to compute representations, providing the flexibility to operate at multiple scales with a computational complexity that scales linearly with image size. This adaptability makes the Swin Transformer particularly well-suited for tasks that require detailed spatial resolution at various scales.
}

\section{Methodology}
\label{Methodology}

The architecture of our developed ODFormer is illustrated in Fig. \ref{fig.architecture}. It consists of three main components:
\begin{itemize}
    \item A multi-scale context aggregator for feature initialization.
    \item An ODFormer encoder for the extraction of hierarchical feature maps from RGB images.
    \item An ODFormer decoder that recursively refines and fuses the multi-scale feature maps to generate the final semantic predictions.
\end{itemize}

\subsection{Multi-Scale Context Aggregator}

To transform the given fundus image $\boldsymbol{I}\in\mathbb{R}^{H\times W \times 3}$ into a sequence of patches suitable for processing with the Transformer, with $H$ and $W$ representing the height and width of the image, respectively, the conventional ViT model directly splits the fundus image into non-overlapping patches using linear projection \cite{dosovitskiy2020image}. 
Nevertheless, the ViT architecture poses a significant challenge concerning the absence of long-range dependencies, primarily due to its limited receptive fields. This limitation holds paramount importance in semantic segmentation tasks. To overcome this limitation, we design an MSCA, which first utilizes a series of $m$ atrous convolutions $\omega _{d}$ ($d\in [1,m]\cap\mathbb{Z}$ representing the atrous rate) to extract multi-scale feature maps, which are then aggregated for feature initialization. As the network's depth increases, the atrous rates progressively become larger, leading to an expanded receptive field. This process can be formulated as follows:
\begin{equation}
\begin{aligned}
\boldsymbol{F}_{}^{I} =\mathrm {LN}\bigg({ \underset{5\times5}{\mathrm {Conv2d}} }\Big(\underset{d\in [1,m]}{\mathrm {ConCat}} \big(\omega _{d} (\boldsymbol{I})\big) \Big)\bigg)\in \mathbb{R}^{\frac{H}{4} \times \frac{W}{4} \times C^I},
\end{aligned}
\end{equation}
where $\mathrm {Conv2d}$ denotes a 2D convolution layer, $\mathrm {ConCat}(\cdot)$ represents a concatenation operation performed in the channel dimension, $\boldsymbol{F}^{I}$ denotes the initialized feature maps, which are subsequently fed into our ODFormer encoder, and $C^I$ represents the channel number of $\boldsymbol{F}^{I}$. Sect. \ref{sec.ablation_study} details the experiments conducted to demonstrate the effectiveness of our proposed MSCA.

\subsection{ODFormer Encoder}
\label{sec.odformer_encoder}
Our ODFormer encoder is designed based on the Swin Transformer to generate hierarchical feature maps.
Our ODFormer encoder contains $k$ stages. 
In each stage, $2n_{i}$ ODFormer blocks are stacked sequentially for feature transformation. The input feature maps $\boldsymbol{F}_{i,j-1}^{S}$ are first processed through a novel multi-head self-attention ($\mathrm{MHSA}$) block that incorporates relative position bias to produce the intermediate feature maps $\boldsymbol{F}_{i,j}^{M}$, which are subsequently processed by a multi-layer perceptron ($\mathrm{MLP}$) to generate the encoded feature maps $\boldsymbol{F}_{i,j}^{S}$, where $\boldsymbol{F}_{i,j-1}^{S}, \boldsymbol{F}_{i,j}^{M}, \boldsymbol{F}_{i,j}^{S}  \in \mathbb{R}^{\frac{H}{2^{i+1} } \times \frac{W}{2^{i+1} }\times (2^{i-1}C^{I}) } (i\in [1,k]\cap\mathbb{Z}, j\in [1,2n_{i}]\cap\mathbb{Z})$.

ViT performs self-attention calculations on the input features in a parallel and uniform manner, which results in the neglect of order and position information within the sequence \cite{dosovitskiy2020image}. In contrast, the Swin Transformer incorporates absolute positional encoding to account for the order of tokens \cite{liu2021swin}. However, this approach introduces unique positional encoding for each patch, compromising invariance and neglecting local relationships and structural information within patches. To address this issue, we introduce an additional convolutional layer to extract a relative position bias map and incorporate a residual connection to ensure stable model training. When performing self-attention calculations, let $\boldsymbol{X}^{I}= \mathrm {LN} (\boldsymbol{F}_{i,j-1}^{S})\in \mathbb{R}^{M^{2} \times d}$ and $\boldsymbol{X}^{O} \in \mathbb{R}^{M^{2} \times d}$ be the input and output matrices, respectively, where $\mathrm {LN}(\cdot )$ denotes layer normalization, $M$ denotes the side length of each patch, and $d$ represents the query/key dimension. We first construct the relative position bias map for each head, as follows:
\begin{equation}
\begin{aligned}
\boldsymbol{B}_{r}=\mathrm{Sigmoid}\big(\underset{3\times3}{\mathrm{Conv2d}}(\boldsymbol{X}^{I})\big)\in \mathbb{R}^{M^{2}\times M^{2}}, \ r\in [1,N_{h}],
\end{aligned}
\end{equation}
where $N_{h}$ denotes the number of heads. Subsequently, the query matrix $\boldsymbol{Q}_{}=\boldsymbol{X}^{I}\boldsymbol{W}_{}^{Q} \in \mathbb{R}^{M^{2}\times d }$, key matrix $\boldsymbol{K}_{}=\boldsymbol{X}^{I}\boldsymbol{W}_{}^{K} \in \mathbb{R}^{M^{2}\times d }$, and value matrix $\boldsymbol{V}_{}=\boldsymbol{X}^{I}\boldsymbol{W}_{}^{V} \in \mathbb{R}^{M^{2}\times d }$ are learned for each head, where $\boldsymbol{W}_{}^{Q,K,V}\in \mathbb{R}^{ d \times d }$ are linear projection matrices. Then $\boldsymbol{Q},\boldsymbol{K},\boldsymbol{V}$ are evenly divided into $N_{h}$ equal parts along the dimension direction, and each part is $\boldsymbol{Q}_{r},\boldsymbol{K}_{r},\boldsymbol{V}_{r} \in \mathbb{R}^{M^{2}\times \frac{d}{N_{h}} }$. A multi-head self-attention is then performed as follows:
\begin{equation}
\begin{aligned}
\boldsymbol{H}_{r} =  \mathrm {Softmax}(\frac{\boldsymbol{Q}_{r} \boldsymbol{K}_{r}^{\top } }{\sqrt{\frac{d}{N_{h}}  } } )\times \boldsymbol{V}_{r}+ \boldsymbol{B}_{r}.
\end{aligned}
\end{equation}
$\boldsymbol{X}^{O}$ can therefore be obtained through:
\begin{equation}
\begin{aligned}
\boldsymbol{X}^{O} = \underset{r\in [1, N_{h}]}{\mathrm {ConCat}}(\boldsymbol{H}_{r}^{})+ \boldsymbol{X}^{I}.
\end{aligned}
\end{equation}
The output feature maps $\boldsymbol{F}_{i,j}^{S}$ can then be yielded through:
\begin{equation}
\begin{aligned}
\boldsymbol{F}_{i,j}^{S}=\mathrm {MLP} \big ( \mathrm {LN}( \boldsymbol{F}_{i,j}^{M})\big) + \boldsymbol{F}_{i,j}^{M}.
\end{aligned}
\end{equation}
To produce hierarchical features, our ODFormer block reduces the spatial resolution of the input features through downsampling:
\begin{equation}
\begin{aligned}
\boldsymbol{F}_{i,0}^{S}=\mathrm{LN}\big(\underset{3\times3}{\mathrm{Conv2d}} (\boldsymbol{F}_{i-1,2n_{i-1}}^{S})\big),\ (i \ne1).
\end{aligned}
\end{equation}
The hierarchical encoder feature maps $\mathcal{F}^E=\{\boldsymbol{F}_{1}^{E}, \dots, \boldsymbol{F}_{k}^{E}\}$, where $\boldsymbol{F}_{i}^{E} = \boldsymbol{F}_{i,2n_{i}}^{S}$, can therefore be extracted.

\begin{figure}[t!]
\includegraphics[width=\columnwidth]{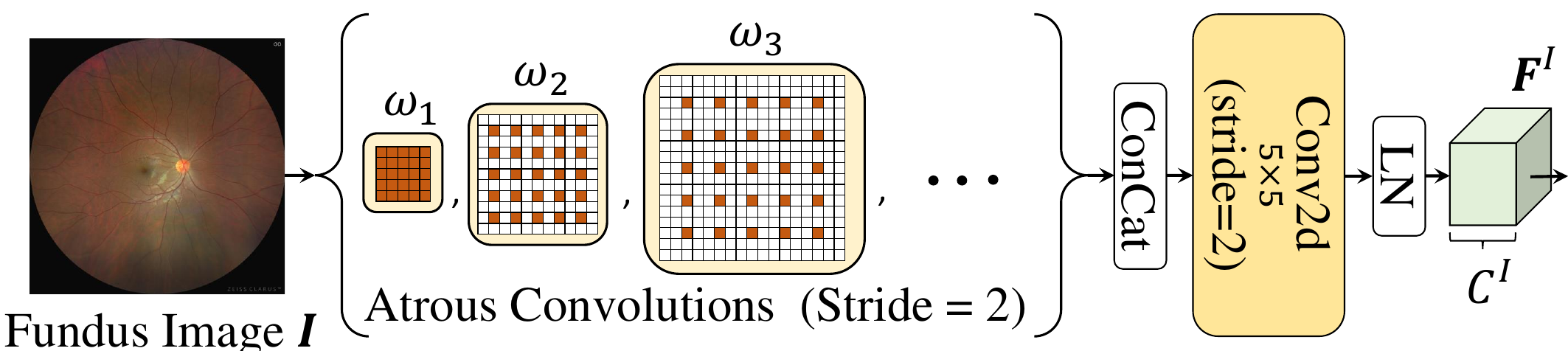}
\caption{The architecture of multi-scale context aggregator.}
\label{fig.MSCA}
\end{figure}

\subsection{ODFormer Decoder}
\begin{figure}[t!]
\includegraphics[width=0.49\textwidth]{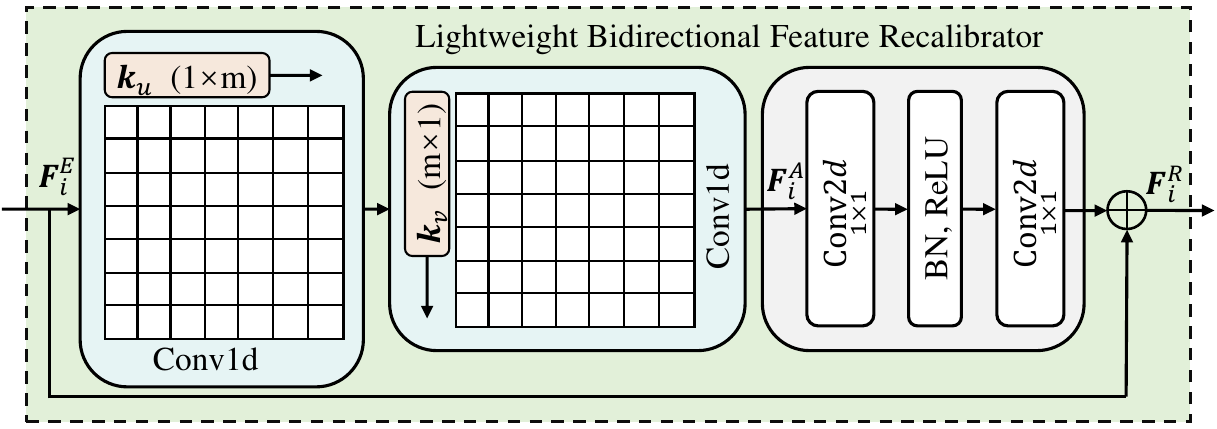}
\caption{The architecture of lightweight bidirectional feature recalibration.}
\label{fig.method_part3}
\end{figure}

We design our ODFormer decoder based on \clr{UPerNet} \cite{xiao2018unified} to recursively recalibrate and fuse the hierarchical feature maps extracted by our ODFormer encoder. However, the use of simplistic upsampling operations, such as bilinear interpolation, can lead to the loss of local information, resulting in unsatisfactory boundary details in semantic segmentation. To overcome this limitation, we aim to incorporate more relevant local information into the features before performing upsampling. To achieve this, we design an LBFR (see Fig. \ref{fig.method_part3}) to capture additional local information while simultaneously reducing computational complexity and the number of network parameters. This approach allows us to enhance the representation of local details in the fundus segmentation process.

Our LBFR first extracts an attention map $\boldsymbol{F}_{i}^{A} \in \mathbb{R}^{\frac{H}{2^{i+1}}\times\frac{W}{2^{i+1}}\times C^{D}}$ from the encoder feature maps $\boldsymbol{F}_{i}^{E}$. In this process, the conventional 2D convolution with a large kernel size is replaced with two 1D spatial separable convolutions, as follows:
\begin{equation}
\begin{aligned}
\boldsymbol{F}_{i}^{A} = 
\boldsymbol{k}_{v} 
\circledast 
(\boldsymbol{k}_{u} 
\circledast \boldsymbol{F}_{i}^{E}),
\end{aligned}
\end{equation}
where $\circledast$ denotes the convolution operation, and $\boldsymbol{k}_{u}$ and $\boldsymbol{k}_{v}$ represent 1D convolutions that operate in the horizontal and vertical directions, respectively. The performance comparison of different convolutional
strategies that possess receptive fields of the same size are presented in Sect. \ref{sec.ablation_study}. The original encoder feature maps $\boldsymbol{F}_{i}^{E}$ are then recalibrated and fused through:
\begin{equation}
\begin{aligned}
\boldsymbol{F}_{i}^{R} = 
\underset{1\times 1}{\mathrm{Conv2d}} \bigg(\mathrm{ReLU}\Big(\mathrm{BN}\big(\underset{1\times 1}{\mathrm{Conv2d}} (\bm{F}_{i}^{A})\big) \Big)\bigg)
+ \boldsymbol{F}_{i}^{E}.
\end{aligned}
\end{equation}
Our ODFormer decoder recursively generates a collection of $k$ decoder feature maps $\mathcal{F}^{D}=\{\boldsymbol{F}_{k}^{D},\dots,\boldsymbol{F}_{1}^{D}\}$, where
\begin{equation}
\begin{aligned}
\boldsymbol{F}_{i}^{D} = \boldsymbol{F}_{i}^{R} + \mathrm{Resize}( \boldsymbol{F}_{i+1}^{D}), \ (i\in [1,k]\cap\mathbb{Z}),
\end{aligned}
\end{equation}
where $\mathrm{Resize}(\cdot)$ denotes the resizing operation via bilinear interpolation. The output feature maps from the ODFormer decoder are obtained through:
\begin{equation}
\begin{aligned}
\boldsymbol{F}_{}^{O} =  \underset{3\times 3}{\mathrm{Conv2d}}  \Big(\underset{i\in [1,k]}
{\mathrm{ConCat}} \big(\mathrm{Resize}(
\boldsymbol{F}_{i}^{D}
)\big)\Big)\in  \mathbb{R}^{\frac{H}{4} \times \frac{W}{4}\times K} ,
\end{aligned}
\end{equation}
where $K$ represents the total number of classes. Finally, $\boldsymbol{F}_{}^{O}$ is passed through a bilinear layer to match the size of the input fundus image $\boldsymbol{I}$ for segmentation.

\begin{figure}[t!]
\includegraphics[width=0.5\textwidth]{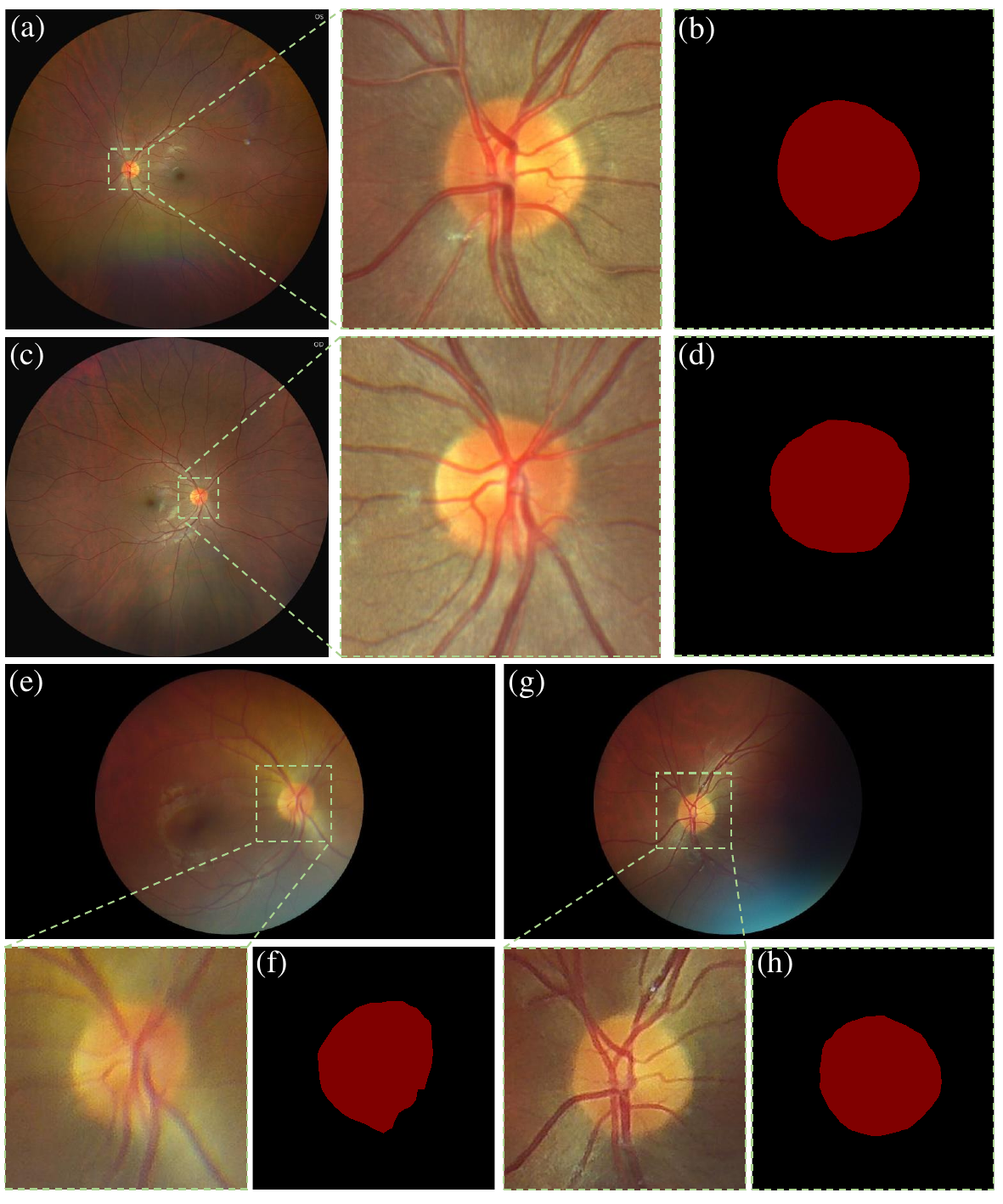}
\caption{Examples of fundus images from a single participant captured using two different cameras: (a)-(d) are the fundus images captured using a Zeiss CLARUS 500 fundus camera and their ground-truth annotations; (e)-(h) 
are the fundus images captured using an NES-1000P handheld mydriasis-free portable fundus camera and their ground-truth annotations.}
\label{fig.samples}
\end{figure}

\begin{figure*}[t!]
\includegraphics[width=0.90\textwidth]{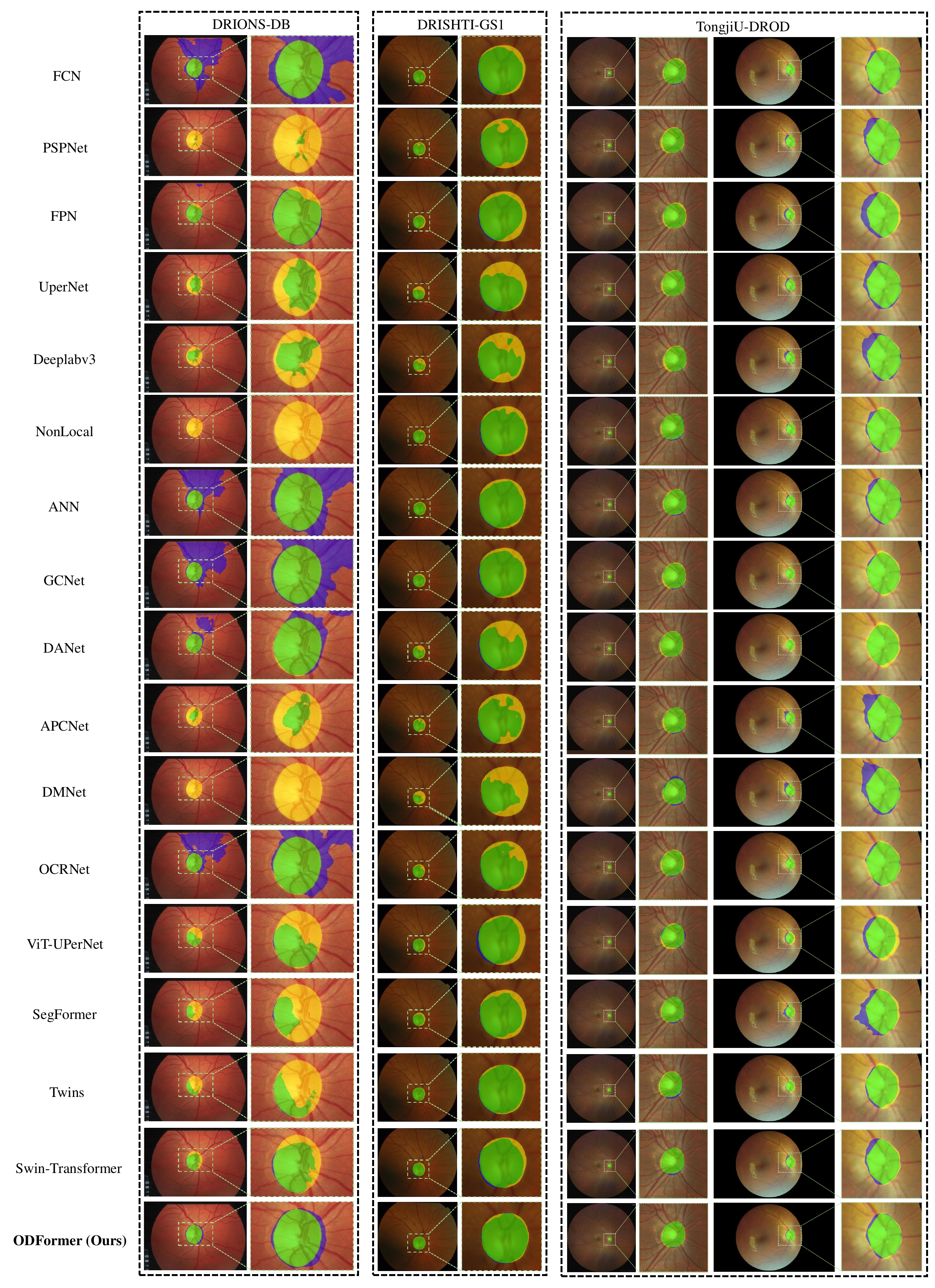}
\caption{
\clr{Qualitative experimental results achieved by ONH detection networks trained on the TongjiU-DROD dataset.}
}
\label{fig.results}
\end{figure*}

\section{Datasets}
\label{sec.datasets}

We first summarize the existing datasets created for ONH detection in Sect. \ref{Existing Datasets}. Then we present a detailed description of our published \clr{TongjiU-DROD} dataset in Sect. \ref{sec.tongji_dataset}.

\subsection{Existing Datasets}
\label{Existing Datasets}

\begin{itemize}
    \item
    
    DRIONS-DB \cite{carmona2008identification} contains 110 pairs of fundus images (resolution: $600 \times 400$ pixels), each with two manually-labeled ground-truth annotations. 90 pairs are used for model training, while the remaining 20 pairs are used for model validation. Within these images, 50 instances exhibit various defects, including illumination artifacts, rim blurredness, and papillary atrophy. These defects could potentially pose challenges to the accurate segmentation of fundus images \cite{abdullah2016localization}.
    
    \item DRISHTI-GS1 \cite{sivaswamy2015comprehensive} contains 101 pairs of high-resolution fundus images (resolution: $2,470 \times 1,760$ pixels) including samples from both healthy and glaucomatous eyes. 81 pairs of images are used for model training, while the remaining 20 pairs of images are used for model validation. For each image in the DRISHTI-GS1 dataset, ground-truth annotations are collected for both ONH and cup regions. These annotations are obtained through the input of four different human experts with varying levels of clinical experience. The average boundaries for both the ONH and cup regions are derived from the manually marked boundaries provided by these experts.  
\end{itemize}

Nonetheless, it is important to note that both these datasets were generated exclusively using a single type of fundus camera. Consequently, semantic segmentation models trained solely on either of these datasets may demonstrate limited generalizability, primarily because of the limited diversity in the data sources. Therefore, one of our primary objectives is to create a diverse fundus image dataset, comprising images captured using various types of fundus cameras. This dataset will facilitate the evaluation of the generalizability of ONH detection networks across different imaging devices.

\subsection{\clr{TongjiU-DROD} Dataset}
\label{sec.tongji_dataset}

\begin{table}[t!]
  \centering
   \settablefont
    \caption{\clr{The segmentation results with respect to different atrous convolutions in MSCA. The best results are shown in bold type.}}
    \label{table.MSCA}
     \begin{tabular}{ c | c c c c }
                  \toprule[1.2pt]

       & Method & IoU (\%) & Acc (\%) & Fsc (\%) \\ 
 \hline
       
               \multirow{9}{*}{{TongjiU-DROD}}& Swin-T  
               & 87.27 & 93.20 & 90.54 \\
               
               & + MSCA ($m=1$)
               & 87.27 & 93.20 & 90.54 \\ 
               
               & + MSCA ($m=2$)
               & 87.53 & 93.35 & \bm{${95.45}$} \\ 

               & + MSCA ($m=3$)
               & 87.81 & 93.51 & 94.06 \\ 

               & + MSCA ($m=4$)
               & 88.05 & 93.64 & 92.53 \\ 
               
               & + MSCA ($m=5$)
               & 88.09 & 93.67 & 91.42 \\ 

               & + MSCA ($m=6$)
               & 88.21 & 93.74 & 93.06 \\ 
               
               & + MSCA ($m=7$)
               & 88.27 & 93.77 & 92.85 \\ 
               
               & + MSCA ($m=8$)
               & \bm{${88.44}$} & \bm{${93.87}$} & 93.19 \\ \hline

                 \multirow{9}{*}{{DRIONS-DB}}& Swin-T  
               & 91.12 & 95.36 & 96.83 \\
               
               & + MSCA ($m=1$)
               & 91.14 & 95.36 & 97.39 \\ 
               
               & + MSCA ($m=2$)
               & 91.14 & 95.36 & 97.39 \\ 

               & + MSCA ($m=3$)
               & 91.37 & 95.49 & \bm{${97.98}$} \\ 

               & + MSCA ($m=4$)
               & 91.43 & 95.52 & 96.83 \\ 
               
               & + MSCA ($m=5$)
               & 91.45 & 95.54 & 96.58 \\ 

               & + MSCA ($m=6$)
               & 91.26 & 95.43 & 97.93 \\ 
               
               & + MSCA ($m=7$)
               & \bm{${91.51}$} & \bm{${95.57}$} & 97.69 \\ 
               
               & + MSCA ($m=8$)
               & 91.30 & 95.45 & 97.96 \\ \hline

                 \multirow{9}{*}{ {DRISHTI-GS}}& Swin-T  
               & 86.28 & 54.81 & 77.15 \\
               
               & + MSCA ($m=1$)
               & 91.99 & 95.83 & 97.32 \\ 
               
               & + MSCA ($m=2$)
               & 92.03 & 95.85 & 97.72 \\ 

               & + MSCA ($m=3$)
               & 92.70 & 96.21 & \bm{${98.29}$} \\ 

               & + MSCA ($m=4$)
               & 93.08 & 96.41 & 92.45 \\ 
               
               & + MSCA ($m=5$)
               & 93.08 & 96.42 & 95.35 \\ 

               & + MSCA ($m=6$)
               & 93.13 & 96.44 & 96.97 \\ 
               
               & + MSCA ($m=7$)
               & 93.60 & \bm{${96.96}$} & 96.69 \\ 
               
               & + MSCA ($m=8$)
               & \bm{${93.79}$} & 96.79 & 97.13 \\ 
               \toprule[1.2pt]
    \end{tabular}
\end{table}

\subsubsection{Data Collection}

The data collection process involves 147 participants, \clr{aged between 18 and 30 years, with an equal distribution of males and females}. All images were collected with the explicit consent of the participants. Data from both eyes were contributed by 53 participants, while 94 participants provided data from only one eye. Each fundus image was captured by two experts using different cameras, yielding two fundus images per eye.

Considering that early indications of disease are often subtle and challenging to discern through direct observation or low-resolution fundus imaging, we first employed a Zeiss CLARUS 500 fundus camera to acquire high-resolution images (resolution: $3,912\times 3,912$ pixels), with the macula at the center, providing a field of view of approximately 133$^\circ$. The Zeiss CLARUS 500 fundus camera is renowned for its true color imaging, high-definition capabilities, and ultra-wide-angle field of view, enabling the capture of high-resolution fundus images with exceptional clarity, achieving resolutions as fine as 7 microns across a range from the macula to the far periphery.

Moreover, fundus images (resolution: $1,920\times 1,088$ pixels), with a field of view of approximately 40$^\circ$, were also captured using an NES-1000P handheld mydriasis-free portable fundus camera provided by Shanghai New Eyes Medical Co., Ltd. Its lightweight and portable design makes it an ideal tool for clinicians who need the flexibility to perform remote examinations \clr{or who do not have access to a desktop fundus camera}.

\begin{table}[t!]\scriptsize
\settablefont
  \begin{center}
  \renewcommand\arraystretch{1.2}
    \caption{\clr{Ablation study to demonstrate the effectiveness of our proposed LBFR. The best results are shown in bold type.}}
    \label{table.LBFR}
     \begin{tabular}{ c | c c c c }
           \toprule[1.25pt]
                       & \multirow{2}*{Method} 
                       & \makebox[0.08\textwidth][c]{TongjiU-DROD} 
                       & \makebox[0.06\textwidth][c]{DRIONS-DB}
                       & \makebox[0.08\textwidth][c]{DRISHTI-GS1}
                       \\
                       \cline{3-5}
       &  &\multicolumn{3}{c}{IoU (\%)}  \\ \hline

                 \multirow{6}{*}{\rotatebox{90}{TongjiU-DROD}}
               & Swin-T 
               & 87.45 & 68.27 & 86.01 \\
               & + strategy (A)
               & 86.75 & \bm{${73.48}$} & \bm{${90.62}$} \\ 
               & + strategy (B)
               & 87.61 & 68.54 & 86.92 \\ 
               & + strategy (C)
               & 86.60 & 72.61 & 88.29 \\ 
               & + LBFR 
               & \bm{${88.22}$} & 70.57 & 86.03 \\ 
               \hline

                 \multirow{6}{*}{\rotatebox{90}{DRIONS-DB}}
                 
               & Swin-T  
               & 53.51 & 90.07 & 83.76 \\
               & + strategy (A)
               & 34.57 & \bm{${90.83}$} & 86.85 \\ 
               & + strategy (B)
               & 54.66 & 90.13 & 86.21 \\ 
               & + strategy (C)
               & 59.70 & 90.35 & 79.98 \\ 
               & + LBFR
               & \bm{${75.48}$} & 90.49 & \bm{${86.91}$} \\ 
               \hline
               
                 \multirow{6}{*}{\rotatebox{90}{DRISHTI-GS1}}& Swin-T  
               & 40.27 & 83.46 & 92.05 \\
               & + strategy (A)
               & 51.69 & 80.97 & 92.86 \\ 
               & + strategy (B)
               & 47.87 & 83.21 & 93.28 \\ 
               & + strategy (C)
               & 50.24 & \bm{${83.55}$} & 92.88 \\ 
               & + LBFR
               & \bm{${59.92}$} & 80.79 & \bm{${93.68}$} \\ 
        \toprule[1.2pt]       
    \end{tabular}
  \end{center}
\end{table}

\begin{table}[t!]
\settablefont
  \begin{center}
  \renewcommand\arraystretch{1.25}
    \caption{\clr{Ablation study to demonstrate the effectiveness of our proposed MSCA and LBFR. The best results are shown in bold type.}}
    \label{table.both}
     \begin{tabular}{ c | c | c | c c c }
     \toprule[1.2pt]
                  & \multirow{2}*{\scriptsize{MSCA}} & \multirow{2}*{\scriptsize{LBFR}} 
                       &\makebox[0.08\textwidth][c]{TongjiU-DROD} 
                       &\makebox[0.06\textwidth][c]{DRIONS-DB}
                       &\makebox[0.08\textwidth][c]{DRISHTI-GS1} 
                  \\
                  \cline{4-6}
      & &  &\multicolumn{3}{c}{IoU (\%)}  \\ \hline
       
                 \multirow{4}{*}{\rotatebox{90}{\scriptsize{TongjiU-DROD}}}& $\times $ & $\times $
               & 87.45 & 68.27 & 86.01 \\
               & $\surd $ & $\times $
               & 87.81 & \bm{${74.27}$} & 87.37 \\ 
               & $\times $ & $\surd $
               & 88.22 & 70.57 & 86.03 \\
               & $\surd $ & $\surd $
               & \bm{${88.35}$} & 74.18 & \bm{${88.39}$} \\ \hline

                 \multirow{4}{*}{\rotatebox{90}{\scriptsize{DRIONS-DB}}}& $\times $ & $\times $
               & 53.51 & 90.47 & 83.76 \\
               & $\surd $ & $\times $
               & 59.66 & \bm{${91.37}$} & 86.05 \\ 
               & $\times $ & $\surd $
               & \bm{${75.48}$} & 90.49 & 86.21 \\
               & $\surd $ & $\surd $
               & 68.52 & 90.53 & \bm{${86.43}$} \\ \hline

                 \multirow{4}{*}{\rotatebox{90}{\scriptsize{DRISHTI-GS1}}}& $\times $ & $\times $
               & 40.27 & 83.46 & 92.05 \\
               & $\surd $ & $\times $
               & 45.05 & 81.69 & 92.70 \\ 
               & $\times $ & $\surd $
               & 50.78 & 81.89 & 92.41 \\
               & $\surd $ & $\surd $
               & \bm{${63.69}$} & \bm{${83.93}$} & \bm{${93.91}$} \\ \hline
               \toprule[1.2pt]

    \end{tabular}

  \end{center}
\end{table}

\begin{table*}[t!]
\settablefont
  \begin{center}
  \renewcommand\arraystretch{1.1}
    \caption{\clr{The ONH detection benchmark, showing quantitative comparisons among 11 SoTA CNNs, five SoTA Transformers, and our proposed ODFormer across three datasets. The best results are shown in bold type}.}
      \label{table.benchmark}
     \begin{tabular}{C{2.5cm} | c | c c c | c c c | c c c}
                \toprule[1.2pt]
                 & \multirow{2}*{Method} & \multicolumn{3}{c}{TongjiU-DROD} & \multicolumn{3}{c}{DRIONS-DB} & \multicolumn{3}{c}{DRISHTI-GS1}  \\ 
                \cline{3-11}
                 & & \makebox[0.05\textwidth][c]{IoU (\%)} 
                 & \makebox[0.05\textwidth][c]{Fsc (\%)} 
                 & \makebox[0.05\textwidth][c]{Acc (\%)} 
                 & \makebox[0.05\textwidth][c]{IoU (\%)} 
                 & \makebox[0.05\textwidth][c]{Fsc (\%)} 
                 & \makebox[0.05\textwidth][c]{Acc (\%)} 
                 & \makebox[0.05\textwidth][c]{IoU (\%)} 
                 & \makebox[0.05\textwidth][c]{Fsc (\%)} 
                 & \makebox[0.05\textwidth][c]{Acc (\%)} 
                 \\ \hline

                  \multirow{17}{*}{{TongjiU-DROD}}
                  & \textbf{FCN} \cite{long2015fully} & 86.16 & 92.56 & 91.91 & 8.83 & 16.23 & \bm{${99.69}$} & 33.26 & 49.92 & 92.87 \\
                  
                  & \textbf{PSPNet} \cite{zhao2017pyramid} 
                  & 86.07 & 92.51 & 90.06 
                  & 31.76 & 48.21 & 33.09 
                  & 66.29 & 79.72 & 67.61 \\
                  
                  & \textbf{FPNNet} \cite{lin2017feature} 
                  & 84.11 & 91.37 & 89.93 
                  & 28.92 & 44.86 & 83.30 
                  & 79.58 & 88.63 & 90.64 \\
                  
                  & \textbf{\clr{UPerNet}} \cite{xiao2018unified} 
                  & 85.67 & 92.28 & 94.26 
                  & 38.82 & 55.93 & 40.42 
                  & 57.45 & 72.97 & 59.53 \\
                  
                  & \textbf{Deeplabv3} \cite{chen2017rethinking} 
                  & 86.65 & 92.85 & 92.33 
                  & 40.35 & 57.50 & 42.56
                  & 67.38 & 80.51 & 70.63 \\ 
                  
                  & \textbf{NonLocal} \cite{wang2018non} 
                  & 87.84 & 93.52 & 93.67 
                  & 29.02 & 44.99 & 29.97 
                  & 62.64 & 77.03 & 63.76 \\
                  
                  & \textbf{ANN} \cite{zhu2019asymmetric} 
                  & 86.88 & 92.98 & 94.33 
                  & 12.59 & 22.36 & 98.94 
                  & 75.02 & 85.73 & 92.27 \\
                  
                  & \textbf{GCNet} 
                  \cite{cao2019gcnet} 
                  & 87.67 & 93.43 & 92.03 
                  & 9.33 & 17.07 & 98.68 
                  & 79.82 & 88.78 & 87.92 \\
                  
                  & \textbf{DANet} 
                  \cite{fu2019dual} 
                  & 87.41 & 93.28 & 90.28 
                  & 10.84 & 19.56 & 95.82 
                  & 81.43 & 89.77 & 84.40 \\            
                  & \textbf{APCNet} \cite{he2019adaptive} 
                  & 85.77 & 92.34 & 91.60 
                  & 37.12 & 54.15 & 38.92 
                  & 69.73 & 82.16 & 72.20 \\
                  
                  & \textbf{DMNet} \cite{he2019dynamic} 
                  & 80.34 & 89.10 & 90.09 
                  & 15.58 & 26.96 & 15.66 
                  & 35.73 & 52.65 & 36.16 \\
                  
                  & \textbf{OCRNet} \cite{yuan2020object}
                  & 87.18 & 93.15 & 91.40 
                  & 20.74 & 34.35 & 86.69 
                  & 80.09 & 89.44 & 82.93 \\

                  & \clr{\textbf{ViT-UPerNet}} 
                  \cite{dosovitskiy2020image} 
                  & \clr{84.13} & \clr{91.38} & \clr{90.70} 
                  & \clr{62.12} & \clr{76.64} & \clr{65.24} 
                  & \clr{74.61} & \clr{85.46} & \clr{77.05}\\

                  & \clr{\textbf{SegFormer}} \cite{xie2021segformer}
                  & \clr{85.35} & \clr{92.10} & \clr{92.59} 
                  & \clr{63.00} & \clr{77.30} & \clr{74.18} 
                  & \clr{78.16} & \clr{87.74} & \clr{80.14}\\
                  
                  & \clr{\textbf{Twins}} 
                  \cite{chu2021twins} 
                  & \clr{87.03} & \clr{93.07} & \clr{91.40} 
                  & \clr{60.00} & \clr{75.00} & \clr{63.78} 
                  & \clr{\bm{${92.16}$}} & \clr{95.92} & \clr{94.46}\\
                  
                  & \textbf{Swin-T} 
                  \cite{liu2021swin} 
                  & 87.45 & 93.31 & \bm{${95.93}$} 
                  & 68.27 & 81.14 & 73.63 
                  & 86.01 & 92.48 & 89.23 \\
                  
                  \cline{2-11}
                  & \textbf{ODFormer (Ours)}  
                  & \bm{${88.35}$} & \bm{${93.82}$}  & 94.37 
                  & \bm{${74.18}$} & \bm{${85.18}$} & 89.63 
                  & 88.39 & \bm{${93.83}$} & \bm{${97.85}$} \\
                   \toprule[1.2pt]

                  \multirow{17}{*}{{DRIONS-DB}}
                  & \textbf{FCN} \cite{long2015fully} & 49.28 & 66.50 & 68.08 & 68.48 & 81.29 & 70.34 & 
                  53.77 & 69.93 & 53.90 \\
                  
                  & \textbf{PSPNet} \cite{zhao2017pyramid} & 60.04 & 75.03 & 81.13 & 90.22 & 94.86 & 96.05 & 85.78 & 92.35 & 89.98 \\
                  
                  & \textbf{FPNNet} \cite{lin2017feature} & 32.35 & 48.89 & 85.98 & 87.92 & 93.57 & 95.08 & 84.39 & 91.53 & 86.33 \\
                  
                  & \textbf{\clr{UPerNet}} \cite{xiao2018unified} & 30.4 & 46.63 & 69.26 & 89.95 & 94.71 & 94.89 & 78.79 & 88.14 & 80.22 \\
                  
                  & \textbf{Deeplabv3} \cite{chen2017rethinking} & 44.93 & 62.00 & 81.17 & 87.01 & 93.05 & 96.37 & 83.76 & 81.16 & 87.97 \\ 
                  
                  & \textbf{NonLocal} \cite{wang2018non} & 65.97 & 79.50 & 76.92 & 89.90 & 94.68 & 96.90 & 84.23 & 91.44 & 86.94 \\  
                  
                  & \textbf{ANN} \cite{zhu2019asymmetric} & 34.62 & 51.43 & 83.07 & 89.82 & 94.64 & 96.52 & 82.26 & 90.26 & 87.61 \\
                  
                  & \textbf{GCNet} \cite{cao2019gcnet} & 55.09 & 71.04 & 89.10 & 88.57 & 93.94 & 98.51 & 81.53 & 89.83 & 86.78 \\
                  
                  & \textbf{DANet} \cite{fu2019dual} & 57.01 & 72.62 & 73.10 & 89.96 & 94.71 & 95.7 & 82.74 & 90.55 & 86.60 \\    
                  
                  & \textbf{APCNet} \cite{he2019adaptive} & 40.79 & 57.94 & 89.20 & 82.51 & 90.42 & \bm{${98.81}$} & 74.70 & 85.52 & 77.95 \\
                  
                  & \textbf{DMNet} \cite{he2019dynamic} & 45.38 & 62.43 & 47.40 & 72.84 & 82.28 & 82.32 & 41.59 & 58.75 & 41.65 \\
                  
                  & \textbf{OCRNet} \cite{yuan2020object} & 35.86 & 52.79 & 91.09 & 87.13 & 93.13 & 97.95 & 75.06 & 85.76 & 79.18 \\

                  & \clr{\textbf{ViT-UPerNet}} 
                  \cite{dosovitskiy2020image} 
                  & \clr{39.76} & \clr{56.89} & \clr{92.68} 
                  & \clr{90.84} & \clr{95.20} & \clr{96.77} 
                  & \clr{75.58} & \clr{86.09} & \clr{86.64}\\

                  & \clr{\textbf{SegFormer}}
                  \cite{xie2021segformer} 
                  & \clr{16.96} & \clr{29.01} & \clr{88.98} 
                  & \clr{90.10} & \clr{94.79} & \clr{96.55} 
                  & \clr{84.04} & \clr{91.33} & \clr{90.20}\\
                  
                  & \clr{\textbf{Twins}} 
                  \cite{chu2021twins} 
                  & \clr{23.88} & \clr{38.55} & \clr{\bm{${94.54}$}} 
                  & \clr{\bm{${91.15}$}} & \clr{\bm{${95.37}$}} & \clr{96.14} 
                  & \clr{85.79} & \clr{92.35} & \clr{87.03}\\                  
            
                  & \textbf{Swin-T} \cite{liu2021swin} & 53.51 & 69.71 & 64.23 & 90.07 & 94.00 & 93.91 & 83.76 & 91.16 & 88.27 \\
                  
                  \cline{2-11}
                  & \textbf{ODFormer (Ours)}  & \bm{${68.52}$} & \bm{${81.32}$} & 80.22 & 90.53 & 95.03 & 96.64 & \bm{${86.43}$} & \bm{${92.72}$} & \bm{${96.80}$} \\
                   \toprule[1.2pt]

                  \multirow{17}{*}{{DRISHTI-GS1}}
                  & \textbf{FCN} 
                  \cite{long2015fully} 
                  & 44.57 & 61.66 & 64.08 
                  & 12.23 & 21.79 & 93.74 
                  & 93.24 & 96.50 & 95.72 \\
                  
                  & \textbf{PSPNet} 
                  \cite{zhao2017pyramid} 
                  & 63.14 & 77.41 & 84.42 
                  & 63.64 & 77.78 & 70.62 
                  & 92.16 & 95.92 & 95.09 \\
                  
                  & \textbf{FPNNet} 
                  \cite{lin2017feature} 
                  & 42.68 & 59.83 & 91.07 
                  & 38.26 & 55.35 & 99.06 
                  & 91.21 & 95.40 & 96.76 \\
                  
                  & \textbf{\clr{UPerNet}}
                  \cite{xiao2018unified} 
                  & 43.73 & 60.85 & 87.81 
                  & 62.75 & 77.11 & 70.03
                  & 92.12 & 95.90 & 96.11 \\
                  
                  & \textbf{Deeplabv3} 
                  \cite{chen2017rethinking} 
                  & 57.35 & 72.90 & 80.07 
                  & 57.02 & 72.63 & 71.29 
                  & 91.90 & 95.78 & 95.25 \\   
                  
                  & \textbf{NonLocal} 
                  \cite{wang2018non} 
                  & 58.98 & 74.30 & 76.23 
                  & 41.90 & 59.05 & 48.61 
                  & 91.92 & 96.32 & 97.03 \\ 
                  
                  & \textbf{ANN} 
                  \cite{zhu2019asymmetric} 
                  & 62.71 & 76.33 & 83.64 
                  & 79.64 & 45.72 & 66.72 
                  & 91.76 & 95.7 & 97.18 \\
                  
                  & \textbf{GCNet} 
                  \cite{cao2019gcnet} 
                  & 58.52 & 73.83 & 80.30 
                  & 65.89 & 79.44 & 76.94 
                  & 92.90 & 96.32 & 95.36 \\
                  
                  & \textbf{DANet} 
                  \cite{fu2019dual} 
                  & 61.48 & 76.14 & 79.58 
                  & 31.86 & 48.33 & \bm{${99.23}$} 
                  & 92.91 & 96.32 & 97.03 \\ 
                  
                  & \textbf{APCNet} 
                  \cite{he2019adaptive} 
                  & 51.85 & 68.29 & 67.40 
                  & 62.78 & 77.13 & 66.98 
                  & 91.93 & 95.80 & 96.20 \\
                  
                  & \textbf{DMNet} 
                  \cite{he2019dynamic} 
                  & 38.06 & 55.16 & 38.11 
                  & 52.80 & 68.11 & 74.62 
                  & 87.91 & 93.57 & 91.52 \\
                  
                  & \textbf{OCRNet} 
                  \cite{yuan2020object} 
                  & 40.47 & 57.62 & \bm{${93.07}$} 
                  & 46.78 & 63.74 & 94.03 
                  & 91.05 & 95.32 & 97.09 \\

                  & \clr{\textbf{ViT-UPerNet}} 
                  \cite{dosovitskiy2020image} 
                  & \clr{43.14} & \clr{60.27} & \clr{90.92} 
                  & \clr{82.58} & \clr{90.46} & \clr{98.61} 
                  & \clr{85.30} & \clr{92.07} & \clr{92.64}\\  

                  & \clr{\textbf{SegFormer}} 
                  \cite{xie2021segformer} 
                  & \clr{35.86} & \clr{52.79} & \clr{91.09} 
                  & \clr{\bm{${87.13}$}} & \clr{\bm{${93.13}$}} & \clr{97.95} 
                  & \clr{75.06} & \clr{85.76} & \clr{79.18}\\     
                  
                  & \clr{\textbf{Twins}} 
                  \cite{chu2021twins} 
                  & \clr{21.65} & \clr{35.59} & \clr{80.89} 
                  & \clr{82.85} & \clr{90.62} & \clr{95.24} 
                  & \clr{92.25} & \clr{95.97} & \clr{\bm{${97.53}$}}\\

                  & \textbf{Swin-T} 
                  \cite{liu2021swin} 
                  & 40.27 & 57.41 & 92.77 
                  & 83.46 & 90.29 & 97.98 
                  & 92.05 & 95.86 & 96.94 \\
                  \cline{2-11}
                  
                  & \textbf{ODFormer (Ours)}  
                  & \bm{${63.69}$} & \bm{${77.81}$} & 88.19 
                  & 83.93 & 91.27 &  96.80 
                  & \bm{${93.91}$} & \bm{${96.86}$} & 96.88  \\
                   \toprule[1.2pt]
    \end{tabular}
  \end{center}
\end{table*}

\subsubsection{Ground Truth Annotation}
\label{subsubsection:Ground truth Collection}
Our TongjiU-DROD dataset was developed in collaboration with three ophthalmologists who were responsible for participant selection, image selection, and diagnosis assignment. Furthermore, two experts independently provided manual ONH ground-truth annotations, which were subsequently reviewed and refined by a senior expert. Fig. \ref{fig.samples} illustrates an example of four fundus images obtained from two eyes belonging to the same participant, along with their respective ground-truth annotations.

\subsubsection{Dataset Preparation}
\label{sec.dataset_preparation}

The data collection process described above yielded a total of 400 pairs of fundus images, among which 360 pairs are used for model training, while the remaining 40 pairs are used for model validation. To avoid potential bias introduced by the similarity of fundus images taken from the same participant, we determined the training and validation sets based on individual participants. This strategy guarantees that fundus images from a given participant do not appear in both the training and validation sets simultaneously.

\section{Experimental Results}
\label{Experiment Results and Discussion}

\subsection{Experimental Setup}
\label{sec.exp_setup}

We employ two publicly available fundus image datasets, DRISHTI-GS1 and DRIONS-DB, along with our created TongjiU-DROD dataset, in our experiments to evaluate the performance of 16 existing SoTA semantic segmentation networks, as well as our proposed ODFormer.

\clr{A square region was first automatically cropped from the input fundus image with a crop size of $512\times 512$ pixels and then resized to $2048\times 512$ pixels for model training. During the training process, several data augmentation strategies, including random horizontal flipping and photometric distortion with a ratio of 0.5, were applied to increase the amount and type of variation of the training set. It should be noted here that although there are various augmentation methods, they should be chosen according to image characteristics and applied carefully with appropriate parameters to increase the reliability and robustness of the methods \cite{goceri2023medical, goceri2023comparison,  goceri2020image}. For fair comparisons across all experiments, we adopt the same training and augmentation strategy in all experiments. All networks are trained for an identical number of iterations, and the best model is selected based on its performance on the validation set. We consistently use the Stochastic Gradient Descent (SGD) optimizer to optimize all models.} Three common evaluation metrics are employed to quantify the model performance: accuracy (Acc), F-score (Fsc), and the Intersection over Union (IoU). 

\clr{Fig. \ref{fig.results} shows the visualization of ONH detection results achieved by models trained on the TongjiU-DROD dataset.
Next, we conduct ablation studies in Sect. \ref{sec.ablation_study} to demonstrate the effectiveness of our novelties. Then, an ONH detection benchmark that provides a detailed performance comparison between the 16 SoTA networks and our ODFormer is presented in Sect. \ref{sec:ONH Detection Benchmark}.} 

\subsection{Ablation Study}
\label{sec.ablation_study}
\clr{We adopt Swin Transformer as the baseline network due to its superior performance across three datasets. To validate the effectiveness of the MSCA structure, 
we test MSCA with $m \in [1,8]$ artous convolutions. As shown in Table \ref{table.MSCA}, the detection performance gradually improves with the increase in the number of atrous convolutions and finally stabilizes when $m=3$. Therefore, in subsequent experiments, we employ MSCA with three atrous convolutions. The results demonstrate that integrating our MSCA with the Swin Transformer leads to significant performance enhancements, achieving improvements of 0.39\%-7.51\% in terms of IoU and 1.15\%-21.14\% in terms of Fsc. }

\clr{Since the LBFR is specifically designed to reduce network parameters while maintaining the same receptive field size, our investigation focuses on various convolutional strategies that possess receptive fields equivalent to the LBFR.}
As shown in Table \ref{table.LBFR}, strategy (A) involves replacing two 1D spatial separable convolutions in LBFR with three $3\times3$ convolutions, while strategy (B) replaces them with one $3\times3$ convolution followed by one $5\times5$ convolution, and strategy (C) replaces them with one $7\times7$ convolution.
The results indicate that our LBFR exhibits superior detection performance and generalizability compared to these strategies. It achieves an improvement of 0.42\% to 1.63\% in terms of IoU, as well as an enhancement in IoU on external datasets ranging from 0.02\% to 21.97\%.
\clr{These improvements underscore the efficacy of our LBFR approach, which also boasts a minimized parameter count, enhancing both efficiency and effectiveness.}

Furthermore, we evaluate the individual effectiveness of MSCA and LBFR, as well as their compatibility. The results shown in Table \ref{table.both} indicate that 
\clr{(1) MSCA alone achieves an improvement of up to 0.36\% in IoU on the same dataset, 
and an IoU increase on external datasets of up to 6.15\%,
(2) LBFR alone results in a maximum IoU improvement of 0.77\% compared to the baseline setup, and an IoU improvement on external datasets of up to 21.97\%, and (3) The combined utilization of MSCA and LBFR modules yields better performance than using these modules independently.}

\subsection{ONH Detection Benchmark}
\label{sec:ONH Detection Benchmark}

\clr{This subsection introduces an ONH detection benchmark, providing both quantitative and qualitative comparisons among 16 SoTA networks and our newly developed ODFormer, all trained on the three aforementioned datasets. Fig. \ref{fig.results} shows examples of the qualitative results, highlighting that ODFormer delivers more accurate and robust results than all other SoTA networks tested on the TongjiU-DROD dataset with the same input. Quantitative comparisons are detailed in Table \ref{table.benchmark}. These results clearly illustrate the superior performance and improved generalizability of our ODFormer over all other networks in ONH detection, emphasizing the effectiveness of Transformer-based architectures. Specifically, compared to the Swin Transformer, ODFormer demonstrates improvements in both Fsc and IoU by approximately 0.51–1.03\% and 0.46–1.86\%, respectively. Furthermore, it is evident that the network's generalizability is significantly improved, with the Fsc seeing increases ranging from 0.98-20.40\%, and IoU improvements spanning 0.47-23.42\%. This underscores ODFormer's robust capability to handle diverse imaging conditions and dataset variations, making it a highly effective tool in ONH detection.}

\section{Discussion}
\clr{
We discuss several limitations in this study and propose potential directions for future work.

We have not incorporated a denoising module into our ODFormer, which could potentially lead to an overfitting problem, thus compromising its generalizability. This limitation might result in poor performance when the model is confronted with new data. As a future work, the performance of the proposed method can be assessed after integrating an effective denoising method since it is known that images can be noisy and denoising should be carefully employed to avoid a loss of information \cite{goceri2023evaluation}.

Moreover, during the feature extraction process, ODFormer does not sufficiently account for the positional relationship between high-level and low-level features, as well as the relative spatial arrangement of encoded features. This can potentially lead to the loss of crucial information regarding relative positions and angles. As another future study, the performance of the ODFormer structure can be compared with the performance of a capsule network-based method because capsule networks can preserve spatial relationships of learned features, and thus, they have been applied for several image classification works \cite{goceri2023classification, goceri2021analysis, goceri2021capsule}.

Additionally, ODFormer can be modified and applied to different medical images, particularly to achieve precise segmentation and classification of nuclei, polyps, and skin lesions, which are essential for accurate diagnosis of cancer types, and diagnosis of several diseases in gynecologic oncology because efficient automated methods are still needed in these fields although there are some methods based on Transformers and several attention mechanisms \cite{goceri2024vision, goceri2024nuclei, idlahcen2024exploring, goceri2021automated, goceri2024polyp}.
}

\section{Conclusion}

\clr{
In this article, we conducted a comprehensive study on semantic segmentation, including establishing a benchmark, developing a novel network based on Swin Transformer, and publishing a dataset specifically for ONH detection. Our experiments validate that our proposed ODFormer achieves SoTA performance for ONH detection compared to all other existing networks. Additionally, our newly published TongjiU-DROD dataset, containing fundus images captured by two distinct cameras, helps alleviate the scarcity of multi-resolution fundus image datasets. This makes it a valuable resource for both research and clinical applications. We believe that the provided benchmark and our proposed dataset will stimulate further research in this area. Furthermore, the techniques we developed can also be applied to solve other general semantic segmentation challenges. In the future, we plan to continue exploring Transformer-based architectures that can optimize feature representations more effectively and efficiently for ONH detection.
}


\begin{thebibliography}{10}
	
	\bibitem{wong2001retinal}
	Tien~Yin Wong, Ronald Klein, Barbara~EK Klein, James~M Tielsch, Larry Hubbard,
	and F~Javier Nieto.
	\newblock Retinal microvascular abnormalities and their relationship with
	hypertension, cardiovascular disease, and mortality.
	\newblock {\em Survey of ophthalmology}, 46(1):59--80, 2001.
	
	\bibitem{sinthanayothin1999automated}
	Chanjira Sinthanayothin, James~F Boyce, Helen~L Cook, and Thomas~H Williamson.
	\newblock Automated localisation of the optic disc, fovea, and retinal blood
	vessels from digital colour fundus images.
	\newblock {\em British journal of ophthalmology}, 83(8):902--910, 1999.
	
	\bibitem{abdullah2016localization}
	Muhammad Abdullah, Muhammad~Moazam Fraz, and Sarah~A Barman.
	\newblock Localization and segmentation of optic disc in retinal images using
	circular hough transform and grow-cut algorithm.
	\newblock {\em PeerJ}, 4:e2003, 2016.
	
	\bibitem{cheng2013superpixel}
	Jun Cheng, Jiang Liu, Yanwu Xu, Fengshou Yin, Damon Wing~Kee Wong, Ngan-Meng
	Tan, Dacheng Tao, Ching-Yu Cheng, Tin Aung, and Tien~Yin Wong.
	\newblock Superpixel classification based optic disc and optic cup segmentation
	for glaucoma screening.
	\newblock {\em IEEE transactions on medical imaging}, 32(6):1019--1032, 2013.
	
	\bibitem{rohrschneider2004determination}
	Klaus Rohrschneider.
	\newblock Determination of the location of the fovea on the fundus.
	\newblock {\em Investigative ophthalmology \& visual science},
	45(9):3257--3258, 2004.
	
	\bibitem{zhang2020intelligent}
	Li~Zhang and Chee~Peng Lim.
	\newblock Intelligent optic disc segmentation using improved particle swarm
	optimization and evolving ensemble models.
	\newblock {\em Applied Soft Computing}, 92:106328, 2020.
	
	\bibitem{fan2023tmi}
	Rui Fan, Christopher Bowd, Nicole Brye, Mark Christopher, Robert~N. Weinreb,
	David~J. Kriegman, and Linda~M. Zangwill.
	\newblock One-vote veto: Semi-supervised learning for low-shot glaucoma
	diagnosis.
	\newblock {\em IEEE Transactions on Medical Imaging}, 42(12):3764--3778, 2023.
	
	\bibitem{li2001automatic}
	Huiqi Li and Opas Chutatape.
	\newblock Automatic location of optic disk in retinal images.
	\newblock In {\em Proceedings 2001 International Conference on Image Processing
		(Cat. No. 01CH37205)}, volume~2, pages 837--840. IEEE, 2001.
	
	\bibitem{fan2022detecting}
	Rui Fan, Christopher Bowd, Mark Christopher, Nicole Brye, James~A Proudfoot,
	Jasmin Rezapour, Akram Belghith, Michael~H Goldbaum, Benton Chuter,
	Christopher~A Girkin, et~al.
	\newblock Detecting glaucoma in the ocular hypertension study using deep
	learning.
	\newblock {\em JAMA ophthalmology}, 140(4):383--391, 2022.
	
	\bibitem{abdel2004detection}
	RA~Abdel-Ghafar, T~Morris, T~Ritchings, and I~Wood.
	\newblock Detection and characterisation of the optic disk in glaucoma and
	diabetic retinopathy.
	\newblock In {\em Proceedings of medical image understanding and analysis},
	pages 20--24, 2004.
	
	\bibitem{fan2023detecting}
	Rui Fan, Kamran Alipour, Christopher Bowd, Mark Christopher, Nicole Brye,
	James~A Proudfoot, Michael~H Goldbaum, Akram Belghith, Christopher~A Girkin,
	Massimo~A Fazio, et~al.
	\newblock Detecting glaucoma from fundus photographs using deep learning
	without convolutions: {Transformer} for improved generalization.
	\newblock {\em Ophthalmology science}, 3(1):100233, 2023.
	
	\bibitem{bharkad2017automatic}
	Sangita Bharkad.
	\newblock Automatic segmentation of optic disk in retinal images.
	\newblock {\em Biomedical Signal Processing and Control}, 31:483--498, 2017.
	
	\bibitem{chrastek2005automated}
	Radim Chr{\'a}stek, Matthias Wolf, Klaus Donath, Heinrich Niemann, Dietrich
	Paulus, Torsten Hothorn, Berthold Lausen, Robert L{\"a}mmer, Christian~Y
	Mardin, and Georg Michelson.
	\newblock Automated segmentation of the optic nerve head for diagnosis of
	glaucoma.
	\newblock {\em Medical image analysis}, 9(4):297--314, 2005.
	
	\bibitem{li2023uniformer}
	Kunchang Li, Yali Wang, Junhao Zhang, Peng Gao, Guanglu Song, Yu~Liu, Hongsheng
	Li, and Yu~Qiao.
	\newblock Uniformer: Unifying convolution and self-attention for visual
	recognition.
	\newblock {\em IEEE Transactions on Pattern Analysis and Machine Intelligence},
	2023.
	
	\bibitem{feng2022polarformer}
	Yaowei Feng, Zhendong Li, Dong Yang, Hongkai Hu, Hui Guo, and Hao Liu.
	\newblock Polarformer: Optic disc and cup segmentation using a hybrid
	cnn-transformer and polar transformation.
	\newblock {\em Applied Sciences}, 13(1):541, 2022.
	
	\bibitem{hussain2023ut}
	Rukhshanda Hussain and Hritam Basak.
	\newblock Ut-net: Combining u-net and transformer for joint optic disc and cup
	segmentation and glaucoma detection.
	\newblock {\em arXiv preprint arXiv:2303.04939}, 2023.
	
	\bibitem{liu2021swin}
	Ze~Liu, Yutong Lin, Yue Cao, Han Hu, Yixuan Wei, Zheng Zhang, Stephen Lin, and
	Baining Guo.
	\newblock Swin transformer: Hierarchical vision transformer using shifted
	windows.
	\newblock In {\em Proceedings of the IEEE/CVF international conference on
		computer vision}, pages 10012--10022, 2021.
	
	\bibitem{xiao2018unified}
	Tete Xiao, Yingcheng Liu, Bolei Zhou, Yuning Jiang, and Jian Sun.
	\newblock Unified perceptual parsing for scene understanding.
	\newblock In {\em Proceedings of the European conference on computer vision
		(ECCV)}, pages 418--434, 2018.
	
	\bibitem{carmona2008identification}
	Enrique~J Carmona, Mariano Rinc{\'o}n, Juli{\'a}n Garc{\'\i}a-Feijo{\'o}, and
	Jos{\'e}~M Mart{\'\i}nez-de-la Casa.
	\newblock Identification of the optic nerve head with genetic algorithms.
	\newblock {\em Artificial intelligence in medicine}, 43(3):243--259, 2008.
	
	\bibitem{sivaswamy2015comprehensive}
	Jayanthi Sivaswamy, S~Krishnadas, Arunava Chakravarty, G~Joshi, A~Syed Tabish,
	et~al.
	\newblock A comprehensive retinal image dataset for the assessment of glaucoma
	from the optic nerve head analysis.
	\newblock {\em JSM Biomedical Imaging Data Papers}, 2(1):1004, 2015.
	
	\bibitem{long2015fully}
	Jonathan Long, Evan Shelhamer, and Trevor Darrell.
	\newblock Fully convolutional networks for semantic segmentation.
	\newblock In {\em Proceedings of the IEEE conference on computer vision and
		pattern recognition}, pages 3431--3440, 2015.
	
	\bibitem{zhao2017pyramid}
	Hengshuang Zhao, Jianping Shi, Xiaojuan Qi, Xiaogang Wang, and Jiaya Jia.
	\newblock Pyramid scene parsing network.
	\newblock In {\em Proceedings of the IEEE conference on computer vision and
		pattern recognition}, pages 2881--2890, 2017.
	
	\bibitem{zhou2016image}
	Hao Zhou, Jun Zhang, Jun Lei, Shuohao Li, and Dan Tu.
	\newblock Image semantic segmentation based on fcn-crf model.
	\newblock In {\em 2016 International Conference on Image, Vision and Computing
		(ICIVC)}, pages 9--14. IEEE, 2016.
	
	\bibitem{lin2017feature}
	Tsung-Yi Lin, Piotr Doll{\'a}r, Ross Girshick, Kaiming He, Bharath Hariharan,
	and Serge Belongie.
	\newblock Feature pyramid networks for object detection.
	\newblock In {\em Proceedings of the IEEE conference on computer vision and
		pattern recognition}, pages 2117--2125, 2017.
	
	\bibitem{florian2017rethinking}
	L-CCGP Florian and Schroff~Hartwig Adam.
	\newblock Rethinking atrous convolution for semantic image segmentation.
	\newblock In {\em Conference on computer vision and pattern recognition (CVPR).
		IEEE/CVF}, volume~6, 2017.
	
	\bibitem{chen2018encoder}
	Liang-Chieh Chen, Yukun Zhu, George Papandreou, Florian Schroff, and Hartwig
	Adam.
	\newblock Encoder-decoder with atrous separable convolution for semantic image
	segmentation.
	\newblock In {\em Proceedings of the European conference on computer vision
		(ECCV)}, pages 801--818, 2018.
	
	\bibitem{wang2018non}
	Xiaolong Wang, Ross Girshick, Abhinav Gupta, and Kaiming He.
	\newblock Non-local neural networks.
	\newblock In {\em Proceedings of the IEEE conference on computer vision and
		pattern recognition}, pages 7794--7803, 2018.
	
	\bibitem{zhu2019asymmetric}
	Zhen Zhu, Mengde Xu, Song Bai, Tengteng Huang, and Xiang Bai.
	\newblock Asymmetric non-local neural networks for semantic segmentation.
	\newblock In {\em Proceedings of the IEEE/CVF international conference on
		computer vision}, pages 593--602, 2019.
	
	\bibitem{cao2019gcnet}
	Yue Cao, Jiarui Xu, Stephen Lin, Fangyun Wei, and Han Hu.
	\newblock Gcnet: Non-local networks meet squeeze-excitation networks and
	beyond.
	\newblock In {\em Proceedings of the IEEE/CVF international conference on
		computer vision workshops}, pages 0--0, 2019.
	
	\bibitem{fu2019dual}
	Jun Fu, Jing Liu, Haijie Tian, Yong Li, Yongjun Bao, Zhiwei Fang, and Hanqing
	Lu.
	\newblock Dual attention network for scene segmentation.
	\newblock In {\em Proceedings of the IEEE/CVF conference on computer vision and
		pattern recognition}, pages 3146--3154, 2019.
	
	\bibitem{he2019dynamic}
	Junjun He, Zhongying Deng, and Yu~Qiao.
	\newblock Dynamic multi-scale filters for semantic segmentation.
	\newblock In {\em Proceedings of the IEEE/CVF International Conference on
		Computer Vision}, pages 3562--3572, 2019.
	
	\bibitem{liu2024playing}
	Chuang-Wei Liu, Qijun Chen, and Rui Fan.
	\newblock Playing to vision foundation model's strengths in stereo matching.
	\newblock {\em arXiv preprint arXiv:2404.06261}, 2024.
	
	\bibitem{han2022survey}
	Kai Han, Yunhe Wang, Hanting Chen, Xinghao Chen, Jianyuan Guo, Zhenhua Liu,
	Yehui Tang, An~Xiao, Chunjing Xu, Yixing Xu, et~al.
	\newblock A survey on vision transformer.
	\newblock {\em IEEE transactions on pattern analysis and machine intelligence},
	45(1):87--110, 2022.
	
	\bibitem{tang2020defusionnet}
	Chang Tang, Xinwang Liu, Xiao Zheng, Wanqing Li, Jian Xiong, Lizhe Wang,
	Albert~Y Zomaya, and Antonella Longo.
	\newblock Defusionnet: Defocus blur detection via recurrently fusing and
	refining discriminative multi-scale deep features.
	\newblock {\em IEEE Transactions on Pattern Analysis and Machine Intelligence},
	44(2):955--968, 2020.
	
	\bibitem{vaswani2017attention}
	Ashish Vaswani, Noam Shazeer, Niki Parmar, Jakob Uszkoreit, Llion Jones,
	Aidan~N Gomez, {\L}ukasz Kaiser, and Illia Polosukhin.
	\newblock Attention is all you need.
	\newblock {\em Advances in neural information processing systems}, 30, 2017.
	
	\bibitem{li2024roadformer}
	Jiahang Li, Yikang Zhan, Peng Yun, Guangliang Zhou, Qijun Chen, and Rui Fan.
	\newblock Roadformer: Duplex transformer for rgb-normal semantic road scene
	parsing.
	\newblock {\em IEEE Transactions on Intelligent Vehicles}, 2024.
	\newblock {DOI}: 10.1109/TIV.2024.3388726.
	
	\bibitem{li2024hapnet}
	Jiahang Li, Peng Yun, Qijun Chen, and Rui Fan.
	\newblock Hapnet: Toward superior rgb-thermal scene parsing via hybrid,
	asymmetric, and progressive heterogeneous feature fusion.
	\newblock {\em arXiv preprint arXiv:2404.03527}, 2024.
	
	\bibitem{strudel2021segmenter}
	Robin Strudel, Ricardo Garcia, Ivan Laptev, and Cordelia Schmid.
	\newblock Segmenter: Transformer for semantic segmentation.
	\newblock In {\em Proceedings of the IEEE/CVF international conference on
		computer vision}, pages 7262--7272, 2021.
	
	\bibitem{yuan2020object}
	Yuhui Yuan, Xilin Chen, and Jingdong Wang.
	\newblock Object-contextual representations for semantic segmentation.
	\newblock In {\em Computer Vision--ECCV 2020: 16th European Conference,
		Glasgow, UK, August 23--28, 2020, Proceedings, Part VI 16}, pages 173--190.
	Springer, 2020.
	
	\bibitem{xie2021segformer}
	Enze Xie, Wenhai Wang, Zhiding Yu, Anima Anandkumar, Jose~M Alvarez, and Ping
	Luo.
	\newblock Segformer: Simple and efficient design for semantic segmentation with
	transformers.
	\newblock {\em Advances in neural information processing systems},
	34:12077--12090, 2021.
	
	\bibitem{chu2021twins}
	Xiangxiang Chu, Zhi Tian, Yuqing Wang, Bo~Zhang, Haibing Ren, Xiaolin Wei,
	Huaxia Xia, and Chunhua Shen.
	\newblock Twins: Revisiting the design of spatial attention in vision
	transformers.
	\newblock {\em Advances in neural information processing systems},
	34:9355--9366, 2021.
	
	\bibitem{dosovitskiy2020image}
	Alexey Dosovitskiy, Lucas Beyer, Alexander Kolesnikov, Dirk Weissenborn,
	Xiaohua Zhai, Thomas Unterthiner, Mostafa Dehghani, Matthias Minderer, Georg
	Heigold, Sylvain Gelly, et~al.
	\newblock An image is worth 16x16 words: Transformers for image recognition at
	scale.
	\newblock {\em arXiv preprint arXiv:2010.11929}, 2020.
	
	\bibitem{chen2017rethinking}
	Liang-Chieh Chen, George Papandreou, Florian Schroff, and Hartwig Adam.
	\newblock Rethinking atrous convolution for semantic image segmentation.
	\newblock {\em arXiv preprint arXiv:1706.05587}, 2017.
	
	\bibitem{he2019adaptive}
	Junjun He, Zhongying Deng, Lei Zhou, Yali Wang, and Yu~Qiao.
	\newblock Adaptive pyramid context network for semantic segmentation.
	\newblock In {\em Proceedings of the IEEE/CVF Conference on Computer Vision and
		Pattern Recognition}, pages 7519--7528, 2019.
	
	\bibitem{goceri2023medical}
	Evgin Goceri.
	\newblock Medical image data augmentation: techniques, comparisons and
	interpretations.
	\newblock {\em Artificial Intelligence Review}, 56(11):12561--12605, 2023.
	
	\bibitem{goceri2023comparison}
	Evgin Goceri.
	\newblock Comparison of the impacts of dermoscopy image augmentation methods on
	skin cancer classification and a new augmentation method with wavelet
	packets.
	\newblock {\em International Journal of Imaging Systems and Technology},
	33(5):1727--1744, 2023.
	
	\bibitem{goceri2020image}
	Evgin Goceri.
	\newblock Image augmentation for deep learning based lesion classification from
	skin images.
	\newblock In {\em 2020 IEEE 4th International conference on image processing,
		applications and systems (IPAS)}, pages 144--148. IEEE, 2020.
	
	\bibitem{goceri2023evaluation}
	Evgin Goceri.
	\newblock Evaluation of denoising techniques to remove speckle and gaussian
	noise from dermoscopy images.
	\newblock {\em Computers in Biology and Medicine}, 152:106474, 2023.
	
	\bibitem{goceri2023classification}
	Evgin Goceri.
	\newblock Classification of skin cancer using adjustable and fully
	convolutional capsule layers.
	\newblock {\em Biomedical Signal Processing and Control}, 85:104949, 2023.
	
	\bibitem{goceri2021analysis}
	Evgin Goceri.
	\newblock Analysis of capsule networks for image classification.
	\newblock In {\em International conference on computer graphics, visualization,
		computer vision and image processing}, 2021.
	
	\bibitem{goceri2021capsule}
	Evgin Goceri.
	\newblock Capsule neural networks in classification of skin lesions.
	\newblock In {\em International conference on computer graphics, visualization,
		computer vision and image processing}, pages 29--36, 2021.
	
	\bibitem{goceri2024vision}
	Evgin Goceri.
	\newblock Vision transformer based classification of gliomas from
	histopathological images.
	\newblock {\em Expert Systems with Applications}, 241:122672, 2024.
	
	\bibitem{goceri2024nuclei}
	Evgin Goceri.
	\newblock Nuclei segmentation using attention aware and adversarial networks.
	\newblock {\em Neurocomputing}, page 127445, 2024.
	
	\bibitem{idlahcen2024exploring}
	Ferdaous Idlahcen, Ali Idri, and Evgin Goceri.
	\newblock Exploring data mining and machine learning in gynecologic oncology.
	\newblock {\em Artificial Intelligence Review}, 57(2):20, 2024.
	
	\bibitem{goceri2021automated}
	Evgin Goceri.
	\newblock Automated skin cancer detection: where we are and the way to the
	future.
	\newblock In {\em 2021 44th international conference on telecommunications and
		signal processing (TSP)}, pages 48--51. IEEE, 2021.
	
	\bibitem{goceri2024polyp}
	Evgin Goceri.
	\newblock Polyp segmentation using a hybrid vision transformer and a hybrid
	loss function.
	\newblock {\em Journal of Imaging Informatics in Medicine}, pages 1--13, 2024.
	
\end{thebibliography}
\end{document}